\documentclass[aps, reprint, longbibliography, noeprint]{revtex4-2}
\usepackage[english]{babel}
\usepackage[utf8]{inputenc}
\usepackage{amsmath}
\usepackage{amsfonts}

\usepackage[colorlinks=true,linkcolor=black,citecolor=black]{hyperref}
\usepackage[capitalise]{cleveref}
\usepackage{url}
\usepackage{natbib}
\usepackage{graphicx}
\graphicspath{ {figures/} }
\usepackage{xspace}
\newcommand{\ket}[1]{\left\vert #1 \right\rangle}
\newcommand{\op}[2]{\vert #1 \rangle\langle #2 \vert}
\newcommand{\tu}[1]{\textup{#1}}
\newcommand{\dd}{\tu{d}}
\newcommand{\Hc}{\tu{H.c.}}
\newcommand{\sgn}{\tu{sgn}}
\newcommand{\T}{\mathcal{T}}
\renewcommand{\vec}[1]{\ensuremath{{\bf{#1}}\xspace}}
\newcommand{\unit}[1]{\ensuremath{\hat{\vec{#1}}}\xspace}

\begin{document}
\date{\today}
\title{Green's function approach to interacting lattice polaritons and optical nonlinearities in subwavelength arrays of quantum emitters}
\author{Simon Panyella Pedersen$ ^* $, Georg M. Bruun$ ^{\dagger} $, and Thomas Pohl$ ^* $ }
\affiliation{ $ ^* $ Institute for Theoretical Physics, TU Wien, Wiedner Hauptstraße 8-10/136, A-1040 Vienna, Austria}
\affiliation{ $ ^{\dagger} $ Department of Physics and Astronomy, Aarhus University, DK-8000 Aarhus C, Denmark}

\begin{abstract}
	Sub-wavelength arrays of quantum emitters offer an efficient free-space approach to coherent light-matter interfacing, using ultracold atoms or two-dimensional solid-state quantum materials. The combination of collectively suppressed photon-losses and emerging optical nonlinearities due to strong photon-coupling to mesoscopic numbers of emitters holds promise for generating nonclassical light and engineering effective interactions between freely propagating photons. While most studies have thus far relied on numerical simulations, we describe here a diagrammatic Green's function approach that permits analytical investigations of nonlinear processes. We illustrate the method by deriving a simple expression for the scattering matrix that describes photon-photon interactions in an extended two-dimensional array of quantum emitters, and reproduces the results of numerical simulations of coherently driven arrays. The approach yields intuitive insights into the nonlinear response of the system and offers a promising framework for a systematic development of a  theory for interacting photons and many-body effects on collective radiance in two-dimensional arrays of quantum emitters. 
\end{abstract}

\maketitle

Recent progress in controlling quantum many-body systems opens up new ways to engineer light-matter interfaces, in which freely propagating photons can be coupled strongly to assemblies of quantum emitters \cite{rui_subradiant_2020,srakaew_subwavelength_2023,kestler_stateinsensitive_2023,lechner_lightmatter_2023,luan_integration_2020,yu_twodimensional_2019,louca_interspecies_2023,datta_highly_2022,stolz_quantumlogic_2022,stiesdal_controlled_2021,chu_independent_2023,zanner_coherent_2022}. For example, the trapping of atoms in optical lattices \cite{gross_quantum_2017} or the confinement of semiconductor excitons in moir\'e lattices of two-dimensional materials \cite{yu_moire_2017,baek_highly_2020,regan_emerging_2022} makes it possible to create regular arrays of saturable quantum emitters with subwavelength lattice spacing. Importantly, the collective interaction of light with such quantum optical metasurfaces allows one to inhibit photon-scattering losses while generating strong mode-selective coupling to an incident light field \cite{bettles_enhanced_2016,asenjo-garcia_exponential_2017,solomons_universal_2024}. This was demonstrated in experiments with ultracold atoms in optical lattices \cite{rui_subradiant_2020} and holds promise for applications, from enhancing the efficiency of optical quantum memories \cite{facchinetti_storing_2016,ruostekoski_arrays_2017,manzoni_optimization_2018,guimond_subradiant_2019},  coherent wavefront shaping \cite{ballantine_optical_2020,ballantine_cooperative_2021,ballantine_optical_2022,bassler_linear_2023}, to the generation and manipulation of nonclassical states of light \cite{moreno-cardoner_quantum_2021,perczel_topological_2017,asenjo-garcia_optical_2019,zhang_photonphoton_2022,shahmoon_cooperative_2017,pedersen_quantum_2023}. 

The system can be described within a Markov approximation that permits integrating out the photonic degrees of freedom and yields an effective Lindblad master equation for the many-body dynamics of the quantum emitters \cite{asenjo-garcia_exponential_2017,solomons_universal_2024,facchinetti_storing_2016,ruostekoski_arrays_2017,manzoni_optimization_2018,guimond_subradiant_2019,ballantine_optical_2020,ballantine_cooperative_2021,ballantine_optical_2022,bassler_linear_2023,moreno-cardoner_quantum_2021,perczel_topological_2017,asenjo-garcia_optical_2019,zhang_photonphoton_2022,shahmoon_cooperative_2017,pedersen_quantum_2023}. The Markovian master equation can be solved numerically for a few incident photons to obtain the optical response of the emitters by recovering the underlying photonic dynamics from an input-output relation for the light field. 

\begin{figure}
	\centering
	\includegraphics[width=\columnwidth]{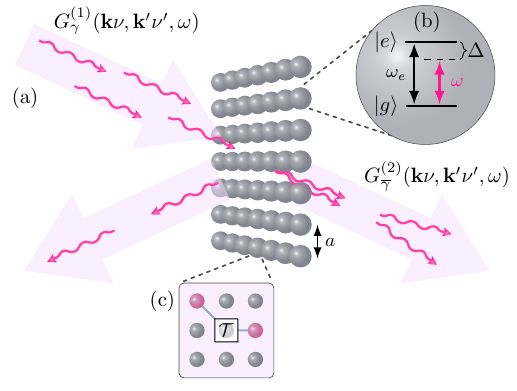}
	\caption{\label{fig:illustration} (a) Illustration of the considered setup, whereby incident photons interact with a 2D array of two-level quantum emitters, depicted in panel (b). Saturation of the two-level emitters generates photon-photon interactions that can be described by (c) a scattering $ \T $-matrix for photon-dressed emitter excitations and a two-photon propagator $ G_{\bar{\gamma}}^{(2)} $. Scattering leads to the correlated transmission of bound photon-pairs, which causes bunching as well as antibunching of the emitted light.}
\end{figure}

In this paper, we discuss a different approach that permits an analytic treatment of emerging photon correlations by considering nonlinear optical processes within a scattering formalism. Here, one employs a polariton picture in terms of dressed photons and dressed excitations that finds broad application in optics, from semiconductor microcavities \cite{carusotto_quantum_2013}, to light-propagation in continuous optical media or discrete chains of emitters \cite{schrinski_polariton_2022}. We will specifically consider a subwavelength square lattice of two-level emitters. Describing the dynamics in terms of time-dependent Green's functions, our method does generally not rely on the Markovian approximation but recovers the results of the master equation in the Markovian limit. Treating the two-level emitters as a lattice of hardcore bosons yields dressed propagators for polaritonic excitations of the array, from which one can obtain effective polariton interactions in terms of a scattering matrix that describes the correlated collective emission of photons. We illustrate the method by deriving analytical and intuitive expressions for the scattering of two incident photons. Moreover, we show that the method yields a simple expression for spatio-temporal photon correlations that agrees remarkably well with the numerical simulation of the effective master equation and reveals the emergence of a two-photon bound state, explaining the specific form of the observed temporal photonic pair-correlations. Our approach thus yields intuitive insights into the leading-order nonlinear response of quantum-emitter arrays and provides a powerful framework for developing a systematic description of many-body effects \cite{fetter_quantum_1971,bruus_manybody_2003,abrikosov_methods_1975} regarding effective photon interactions and collective radiance in extended arrays of quantum emitters.

\section{The System}
\Cref{fig:illustration} illustrates the considered system, which consists of $ N $ two-level emitters that are arranged in a two-dimensional array with a lattice spacing $ a $ in the $ xy $-plane at $ z = 0 $. The system is  described by the Hamiltonian 
\begin{align}\label{eq:realspaceH}
	\begin{split}
		H =& \sum_{n}\omega_{e}\hat{\sigma}_{n}^{\dagger}\hat{\sigma}_{n} + \sum_{\vec{k}, \nu}\omega_{\vec{k}}\hat{b}_{\vec{k}\nu}^{\dagger}\hat{b}_{\vec{k}\nu}\\
		& + \sum_{n, \vec{k}, \nu}\left(\bar{g}_{\vec{k}\nu}e^{-i\vec{k}\cdot\vec{r}_{n}}\hat{b}_{\vec{k}\nu}^{\dagger}\hat{\sigma}_{n} + \Hc\right),
	\end{split}
\end{align}
where the raising and lowering operators, $ \hat{\sigma}^{\dagger}_{n} = \op{e_{n}}{g_{n}} $ and $ \hat{\sigma}_{n} $, are defined by the ground state $ \ket{g_{n}} $ and excited state $ \ket{e_{n}} $ of the $ n $'th emitter, and $ \omega_{e} = 2\pi c/\lambda_{e} $ denotes the corresponding transition frequency ($ c $ is the speed of light). While the first term, thus, describes the isolated dynamics of the two-level emitters, the second term captures the bare propagation of photons with the linear vacuum dispersion $ \omega_{\vec{k}} = c|\vec{k}| $. The corresponding operator $ \hat{b}_{\vec{k}\nu} $ annihilates photons with momentum $ \vec{k} $ and unit polarization vector $ \unit{e}_{\vec{k}\nu} $, labeled by the index $ \nu $. The third term in \cref{eq:realspaceH} accounts for the dipole coupling between emitters and photons, with a coupling strength $ \bar{g}_{\vec{k}\nu} = i\sqrt{\omega_{\vec{k}}/2\hbar\epsilon_{0}V}\unit{e}_{\vec{k}\nu}^{\dagger}\vec{d} $, where $ V $ is a quantization volume for the photonic modes. To be specific, we assume here that the two-level emitters couple to right-circularly polarized light in the $ xy $-plane of the array, such that the corresponding transition dipole moment can be written as $ \vec{d} = d(1, i, 0)^{T}/\sqrt{2} $.

This system can be described within an input-output formalism \cite{asenjo-garcia_exponential_2017,solomons_universal_2024,facchinetti_storing_2016,ruostekoski_arrays_2017,manzoni_optimization_2018,guimond_subradiant_2019,ballantine_optical_2020,ballantine_cooperative_2021,ballantine_optical_2022,bassler_linear_2023,moreno-cardoner_quantum_2021,perczel_topological_2017,asenjo-garcia_optical_2019,zhang_photonphoton_2022,shahmoon_cooperative_2017,pedersen_quantum_2023} that focuses on the spin-dynamics of the emitters. Here, one integrates out the photonic degrees of freedom and applies a Markov approximation to express the total light field as
\begin{align}\label{eq:in_out}
	\hat{\vec{E}}(\vec{r}, t) = \hat{\vec{E}}^{(0)}(\vec{r}, t) + \mu_{0}\omega_{e}^2\sum_{n}\vec{G}(\vec{r}, \vec{r}', \omega_{e})\vec{d}\hat{\sigma}_{n}(t),
\end{align}
where $ \hat{\vec{E}}^{(0)}(\vec{r}, t) = \frac{i}{\sqrt{2\hbar\epsilon_{0}V}}\sum_{\vec{k}, \nu}\sqrt{\omega_{\vec{k}}}\unit{e}_{\vec{k}\nu}e^{i\vec{k}\cdot\vec{r}}\hat{b}_{\vec{k}\nu}^{(0)} $ is the bare electric field, expressed via the bare photon field, $ \hat{b}_{\vec{k}\nu}^{(0)} $, in the absence of the emitters and the induced light field that is generated by them. The latter is determined by the dyadic Green's function, $ \vec{G}(\vec{r}, \vec{r}', \omega_{e}) $ of the free-space electromagnetic field \cite{lehmberg_radiation_1970}. This leads to a master equation for the emitters, with an effective Hamiltonian
\begin{align}\label{eq:Heff}
	\begin{split}
		\hat{H} =& \sum_{n}\omega_{e}\hat{\sigma}_{n}^{\dagger}\hat{\sigma}_{n} + \sum_{n, \vec{k}, \nu}\left(\bar{g}_{\vec{k}\nu}^{*}e^{i\vec{k}\cdot\vec{r}_{n}}\hat{b}_{\vec{k}\nu}^{(0)}\hat{\sigma}_{n}^{\dagger} + \Hc\right)\\
		& - \sum_{n\neq m}J_{mn}\hat{\sigma}_{m}^{\dagger}\hat{\sigma}_{n},
	\end{split}
\end{align}
and Lindbladian
\begin{align}\label{eq:Leff}
	\hat{\mathcal{L}}(\hat{\rho}) = \sum_{n, m}\Gamma_{mn}\left(2\hat{\sigma}_{m} \hat{\rho}\hat{\sigma}^{\dagger}_{n} - \left\{\hat{\sigma}^{\dagger}_{m}\hat{\sigma}_{n}, \hat{\rho}\right\}\right)
\end{align}
that describes the driven-dissipative dynamics of the density matrix $ \hat{\rho} $ for the $ N $-emitter system. The two-body terms result from photon-exchange between the emitters and are determined by $ \frac{\mu_{0}\omega_{e}^2}{\hbar}\vec{d}^{\dagger} \vec{G}(\vec{r}_{m},\vec{r}_{n},\omega_{e})\vec{d} = J_{mn} + i\Gamma_{mn} $ \cite{lehmberg_radiation_1970}. This readily yields the linear optical response captured by the collective Lamb shift
\begin{align}\label{eq:collDelta}
	\tilde{\Delta}_{\vec{k}_{\perp}} = -N^{-1}\sum_{m \neq n}e^{i\vec{k}\cdot(\vec{r}_{m} - \vec{r}_{n})}J_{mn},
\end{align}
and collective linewidth
\begin{align}\label{eq:collGamma}
	\tilde{\Gamma}_{\vec{k}_{\perp}} = N^{-1}\sum_{m, n}e^{i\vec{k}\cdot(\vec{r}_{m} - \vec{r}_{n})}\Gamma_{mn}
\end{align}
of the system, which only depend on the two-dimensional transverse momentum, $ \vec{k}_{\perp} = (k_{x}, k_{y}) $, since $ z = 0 $ in the plane of the array.

Here, we develop a different approach to analyze the system, which describes the nonlinear optical response in terms of quasiparticle scattering processes. To this end, we treat the excited emitter states with bosonic operators, $ \hat{a}_{n} = \hat{\sigma}_{n} $, and introduce a repulsive onsite interaction term $	\frac{U}{2}\sum_{n}\hat{a}_{n}^{\dagger}\hat{a}_{n}^{\dagger}\hat{a}_{n}\hat{a}_{n} $, which prevents double-excitation at a given site as $ U \rightarrow \infty $ \cite{longo_fewphoton_2010,roy_correlated_2011,zheng_persistent_2013,poshakinskiy_biexcitonmediated_2016,ke_inelastic_2019}. Notice that upon taking this limit the two-level nature of the emitters is exactly described, i.e. the employment of bosonic excitations with an infinite on-site repulsion invokes no additional approximation. Next, we define bosonic momentum-space operators $ \hat{a}_{\vec{k}_{\perp}} = a\sum_{n}e^{-i\vec{k}\cdot\vec{r}_{n}}\hat{a}_{n} $ for the emitters. Considering large lattices, one may take the limit $ N \rightarrow \infty $ and $ V \rightarrow \infty $ such that the Hamiltonian can be written as
\begin{align}\label{eq:momspaceH}
	\begin{split}
		H = & \sum_{\nu}\int\frac{\dd^3k}{(2\pi)^3}\omega_{\vec{k}}\hat{b}_{\vec{k}\nu}^{\dagger}\hat{b}_{\vec{k}\nu} + \int_{\tu{BZ}}\frac{\dd^2k_{\perp}}{(2\pi)^2}\omega_{e}\hat{a}_{\vec{k}_{\perp}}^{\dagger}\hat{a}_{\vec{k}_{\perp}}\\
		& - \sum_{\nu}\int\frac{\dd^3k}{(2\pi)^3}\left(g_{\vec{k}\nu}\hat{b}_{\vec{k}\nu}^{\dagger}\hat{a}_{\vec{k}_{\perp}} + \Hc\right)\\
		& + \frac{Ua^2}{2}\int_{\tu{BZ}}\frac{\dd^2k_{\perp}\dd^2k_{\perp}'\dd^2q_{\perp}}{(2\pi)^6}\hat{a}_{\vec{k}_{\perp} + \vec{q}_{\perp}}^{\dagger}\hat{a}_{\vec{k}_{\perp}' - \vec{q}_{\perp}}^{\dagger}\hat{a}_{\vec{k}_{\perp}}\hat{a}_{\vec{k}_{\perp}'},
	\end{split}
\end{align}
where $ g_{\vec{k}\nu} = \sqrt{\omega_{\vec{k}}/2\hbar\epsilon_{0}a^2}\unit{e}_{\vec{k}\nu}^{\dagger}\vec{d} $. Note that the two-dimensional quasi-momentum of the excitations in the discrete array only takes on values within the first Brillouin zone (BZ) of the lattice, and $ \hat{a}_{\vec{k}_{\perp} + \vec{q}_{m}} = \hat{a}_{\vec{k}_{\perp}} $ for any reciprocal lattice vector $ \vec{q}_{m} $. As expressed by the third term in \cref{eq:momspaceH}, identical modes of excited emitters can, hence, couple to different photonic modes, whose momenta differ by a reciprocal lattice vector in the $ xy $-plane. This coupling leads to light scattering and photon losses out of the incident mode (Bragg scattering), but can be suppressed entirely in subwavelength lattices. The last term in \cref{eq:momspaceH}, describes the effective onsite repulsion between emitter excitations. As this interaction is local and uniform it results in a momentum-independent scattering process that exchanges momenta between two modes, while conserving their total momentum. Below, we will exploit this simple form of the effective interaction to derive a compact description of the optical nonlinearity of the array that emerges from saturation of the quantum emitters. We note, however, that the formalism presented here equally applies to other systems and can be straightforwardly used for extended range interactions such as dipolar or van der Waals interactions that appear in atomic \cite{murray_quantum_2016,firstenberg_nonlinear_2016} or solid-state \cite{kazimierczuk_giant_2014,walther_giant_2018,gu_enhanced_2021,datta_highly_2022,orfanakis_rydberg_2022} systems.

\begin{figure}
	\centering
	\includegraphics[width=\columnwidth]{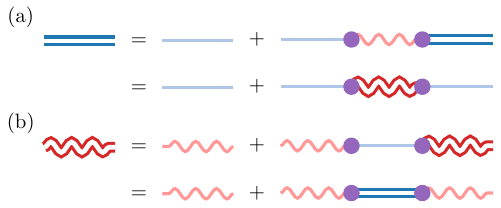}
	\caption{\label{fig:diagrams1} Depiction of the Dyson equation for (a) dressed emitter excitations, and (b) dressed photons. Single particle propagators are depicted as a single wavy or straight line for bare photons or bare excitations, while double lines indicate the dressed propagators due to the emitter-photon coupling (circular vertex). The relations can be recast as, respectively, shown in the second line.}
\end{figure}
\section{Single-photon dynamics}
Let us proceed by considering the single-excitation sector of \cref{eq:momspaceH}, and illustrate the method by re-deriving the known linear optical response of the array. Specifically, we will recover the collective energy shift and decay rate, \cref{eq:collDelta,eq:collGamma}, in the Markovian limit, and show how the dressed-photon propagator recovers the structure of the input-output relation given by \cref{eq:in_out}.

From the first line in the momentum space Hamiltonian, \cref{eq:momspaceH}, we can read off the bare propagator of the emitter excitations
\begin{align}\label{eq:x_bare}
	\begin{split}
		G^{(1)}_{x}(\vec{k}_{\perp}, \vec{k}_{\perp}', \omega)& = (2\pi)^2\delta(\vec{k}_{\perp} - \vec{k}_{\perp}')\mathcal{G}^{(1)}_{x}(\omega),\\
		\mathcal{G}^{(1)}_{x}(\omega)& = \frac{1}{\omega-\omega_{e} + i\eta},
	\end{split}
\end{align}
as well as the bare photon propagator
\begin{align}\label{eq:phot_bare}
	\begin{split}
		G^{(1)}_{\gamma}(\vec{k}, \vec{k}', \omega)& = (2\pi)^3\delta(\vec{k} - \vec{k}')\delta_{\nu, \nu'}\mathcal{G}^{(1)}_{\gamma}(\vec{k},\omega),\\
		\mathcal{G}^{(1)}_{\gamma}(\vec{k},\omega)& = \frac{1}{\omega-\omega_{\vec{k}} + i\eta},
	\end{split}
\end{align}
where $ \eta $ is a positive infinitesimal. The propagators, or Green's functions, of the bare excitations and photons are represented by a straight and wavy line, respectively, in the diagrams shown in \cref{fig:diagrams1}. The light-matter coupling in the second line of \cref{eq:momspaceH} leads to a hybridization and yields single-particle propagators of dressed bosons corresponding to polaritons. The Green's functions of the corresponding quasiparticles are presented as straight (dressed excitations) and wavy (dressed photons) double lines in \cref{fig:diagrams1}. The propagator for the dressed emitter excitations is obtained from the Dyson equation (\cref{fig:diagrams1}a) 
\begin{align}\label{eq:GdressedX}
	\begin{split}
		G^{(1)}_{\bar{x}}(\vec{k}_{\perp}, \vec{k}_{\perp}', \omega) &= (2\pi)^2\delta(\vec{k}_{\perp} - \vec{k}_{\perp}')\mathcal{G}^{(1)}_{\bar{x}}(\vec{k}_{\perp}, \omega)\\
		\mathcal{G}^{(1)}_{\bar{x}}(\vec{k}_{\perp}, \omega) &= \frac{1}{\omega - \omega_{a} - \Sigma(\vec{k}_{\perp}, \omega) + i\eta}.
	\end{split}
\end{align}
Using the expressions for the bare photon propagator $ \mathcal{G}_{\gamma}^{(1)} $ and the light-matter coupling strength, $ g_{\vec{k}\nu} $, we can obtain the self-energy illustrated in \cref{fig:diagrams1}a as 
\begin{align}\label{eq:selfE_x}
	\Sigma(\vec{k}_{\perp},\omega) = \frac{1}{4\pi\hbar\epsilon_{0}a^2}\sum_{\vec{q}_{m}}\int\dd k_{z}\frac{\omega_{\vec{k}}\vec{d}^{\dagger}\vec{Q}\vec{d}}{\omega - \omega_{\vec{k}} + i\eta}.
\end{align}
The projector $ \vec{Q} = 1 - \unit{k}\unit{k}^{\dagger} $ into the plane of photon polarization vectors also defines the dyadic Green's function, $ \vec{G} $, of the free-space electromagnetic field \cite{manzoni_optimization_2018}, which reveals the close connection to the input-output formalism, discussed in the preceding section. Indeed, one can rewrite \cref{eq:selfE_x} as (see \cref{app:selfenergy}) 
\begin{align}\label{eq:selfE_x2}
	\Sigma(\vec{k}_{\perp}, \omega) = \frac{\mu_{0}\omega_{e}^2}{\hbar}N^{-1}\sum_{m, n}e^{i\vec{k}\cdot(\vec{r}_{m} - \vec{r}_{n})}\vec{d}^{\dagger} \vec{G}(\vec{r}_{m}, \vec{r}_{n}, \omega)\vec{d}.
\end{align}
By comparing with \cref{eq:collDelta,eq:collGamma}, we see that the self-energy recovers the collective Lamb shift and decay rate of the array $ \Sigma(\vec{k}_{\perp}, \omega_{e}) = \tilde{\Delta}_{\vec{k}_{\perp}} - i\tilde{\Gamma}_{\vec{k}_{\perp}} $.

The photon propagator cannot be readily calculated analogously, since the summation over reciprocal lattice vectors for the light-matter interaction complicates the direct solution of the Dyson equation shown \cref{fig:diagrams1}b. However, diagrams in the expansion of the Dyson equation can be summed differently to recast the equation as shown in the second line of \cref{fig:diagrams1}b. This yields the dressed-photon propagator
\begin{widetext}
	\begin{align}\label{eq:GdressedPhot}
		G^{(1)}_{\bar{\gamma}}(\vec{k}\nu, \vec{k}' \nu', \omega) = G^{(1)}_{\gamma}(\vec{k}\nu, \vec{k}'\nu', \omega) + (2\pi)^2 \delta_{BZ}(\vec{k}_{\perp}'-\vec{k}_{\perp})\mathcal{G}^{(1)}_{\gamma}(\vec{k},\omega)g_{\vec{k}\nu} \mathcal{G}^{(1)}_{\bar{x}}(\vec{k}'_{\perp},\omega)g_{\vec{k}'\nu'}^{*}\mathcal{G}^{(1)}_{\gamma}(\vec{k}',\omega)
	\end{align}
\end{widetext}
in terms of the Green's function of the dressed emitter excitations, obtained from \cref{eq:GdressedX}. Here
\begin{align}
	\delta_{BZ}(\vec{k}_{\perp} - \vec{k}_{\perp}') = \sum_{\vec{q}_{m}}\delta(\vec{k}_{\perp} - \vec{k}_{\perp}' + \vec{q}_{m})
\end{align}
denotes the delta function for two-dimensional transverse momenta up to a reciprocal lattice vector, $ \vec{q}_{m} $, of the emitter array. This shows how the light emitted from the array, as described by the second term of \cref{eq:GdressedPhot}, consists of Bragg scattered contributions (including both propagating and evanescent waves).

\Cref{eq:GdressedPhot} for the dressed photon propagator resembles the input-output relation \cref{eq:in_out}, as it is given by the bare propagator in the absence of emitters, plus a matter contribution that describes the emission of light by the photon-dressed emitters. The factors of the matter contribution are easily understood as describing the propagation of a free photon that is absorbed in the array, followed by the propagation of the corresponding excitation polariton, and finally the decay of this excitation into a Bragg-scattered freely-propagating photon. The set of \cref{eq:x_bare,eq:phot_bare,eq:GdressedX,eq:selfE_x,eq:GdressedPhot} reproduces the linear optical response of the system given by \cref{eq:in_out,eq:Heff,eq:Leff} obtained from the master equation using the Markovian approximation, which in the present formalism corresponds to setting $ \omega = \omega_{e} $ in the self-energy. In this limit, our diagrammatic theory in principle recovers any result regarding the linear dynamics of the system. For example, the known expressions for the transmission and reflection amplitudes can be found by propagating an initial single-photon plane wave state using the single-photon Green's function \cref{eq:GdressedPhot} (see \cref{app:transrefl}).

\begin{figure}
	\centering
	\includegraphics[width=\columnwidth]{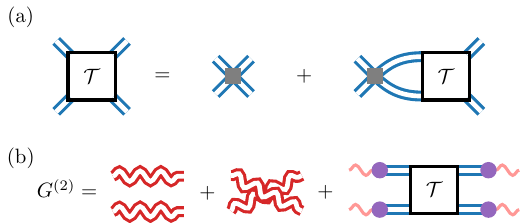}
	\caption{\label{fig:diagrams2} (a) The Bethe-Salpeter equation for the $ \T $-matrix that describes interactions between dressed excitations due to the emitter saturation, as indicated by the square vertex. The two-photon propagator is depicted in panel (b).}
\end{figure}
\section{Photon-photon interactions}
\begin{figure}
	\centering
	\includegraphics[width=\columnwidth]{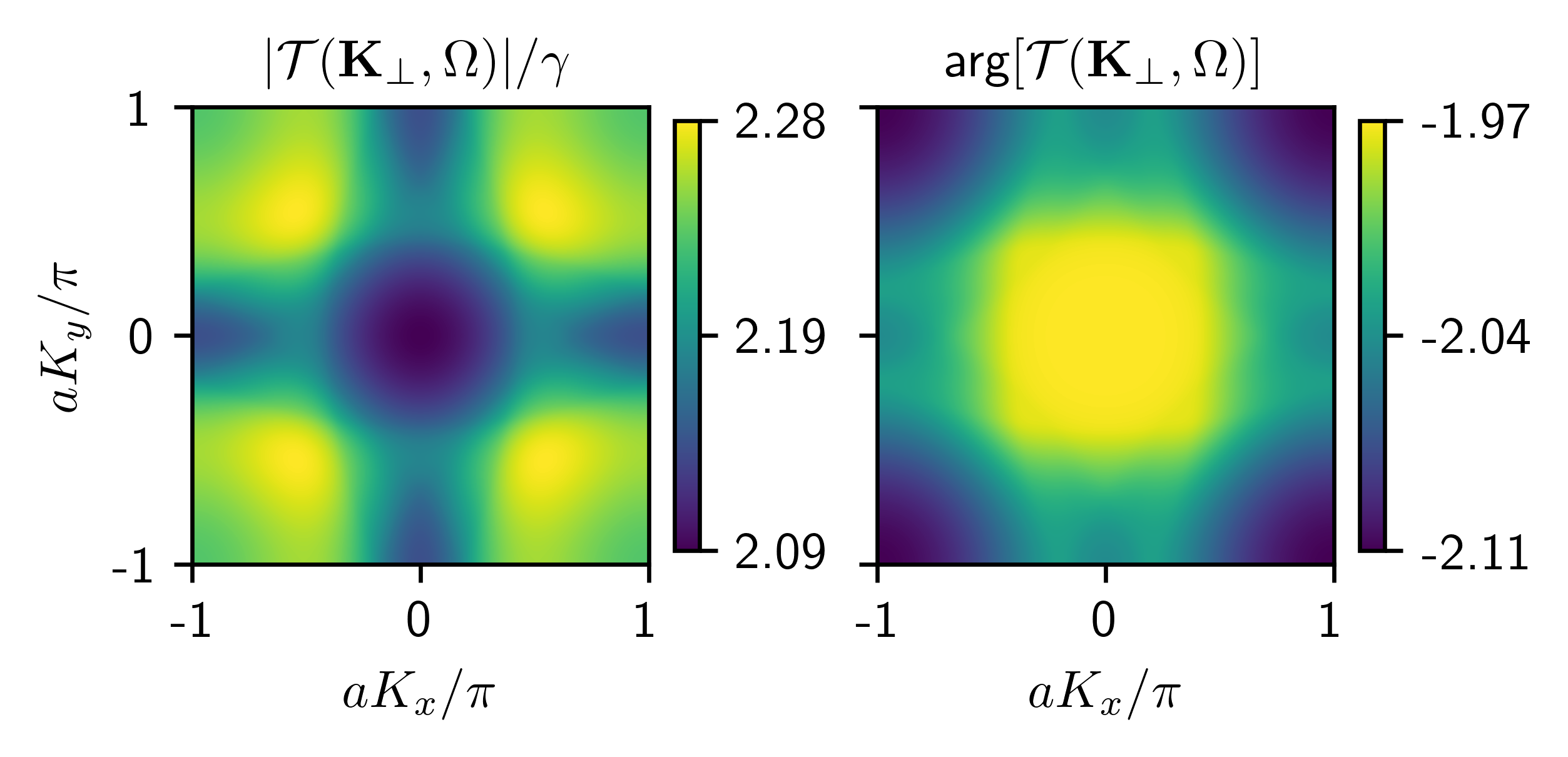}
	\caption{\label{fig:T_matrix} The absolute value and phase of $ \T(\vec{K}_{\perp}, \Omega) $ is plotted as a function of momentum, with the frequency set to the single-photon reflection resonance $ \Omega = 2(\omega_{e} + \tilde{\Delta}_{\vec{0}}) $, and the lattice spacing $ a = 0.6\lambda_{e} $.}
\end{figure}

We now proceed to show how our formalism can describe non-linear effects quite naturally in terms of the scattering of dressed emitters. The Bethe-Salpeter equation for the scattering matrix of two dressed emitters is shown in \cref{fig:diagrams2}a, where only ladder diagrams are non-zero since we consider only two photons \cite{fetter_quantum_1971}. As expressed in \cref{eq:momspaceH}, the local onsite interaction does not carry a dependence on the momenta of the interacting excitations, which makes it possible to solve the Bethe-Salpeter equation algebraically as $ \T = U/[1 - U\Pi] $. Taking the limit $ U \rightarrow \infty $ to describe that maximally one excitation is possible per emitter, returning to the regime of two-level emitters, then yields 
\begin{align}\label{eq:Tsol}
	\T(\vec{K}_{\perp}, \Omega) = -\frac{1}{\Pi(\vec{K}_{\perp}, \Omega)},
\end{align}
where 
\begin{align}\label{eq:pairProp}
	\begin{split}
		\Pi(\vec{K}_{\perp}, \Omega) &= ia^2\int\frac{\dd\omega}{2\pi}\int_{\tu{BZ}}\!\frac{\dd^2q_{\perp}}{(2\pi)^2}\mathcal{G}_{\bar{x}}^{(1)}(\vec{q}_{\perp},\omega)\\
		&\hspace{50pt} \times \mathcal{G}_{\bar{x}}^{(1)}(\vec{K}_{\perp}-\vec{q}_{\perp},\Omega-\omega)
	\end{split}
\end{align}
is the pair propagator for two dressed excitations with total momentum $ \vec{K}_{\perp} = \vec{k}_{\perp,1} + \vec{k}_{\perp,2} $ and total energy $ \Omega = \omega_{1} + \omega_{2} $. Taking $ \Sigma(\vec{k}_{\perp},\omega) \approx \Sigma(\vec{k}_{\perp},\omega_{e}) $ (corresponding to the Markov approximation in a master equation approach) allows one to evaluate the frequency integral analytically, which yields the simple and intuitive expression 
\begin{align}\label{eq:pairPropReduced}
	\Pi(\vec{K}_{\perp}, \Omega) = a^2\!\!\int_{\tu{BZ}}\!\frac{\dd^2q_{\perp}}{(2\pi)^2}\frac{1}{\Omega - 2\omega_{e} - \Sigma(\vec{q}_{\perp}) - \Sigma(\vec{K}_{\perp} - \vec{q}_{\perp})}
\end{align}
for the pair propagator. With this relation and using the self-energy \cref{eq:selfE_x}, the integral in \cref{eq:Tsol} is readily evaluated numerically. \Cref{fig:T_matrix} shows the resulting scattering matrix for a lattice spacing of $ a = 0.6\lambda_{e} $ and incident light on the reflection resonance of the array (in the figure $ \gamma = d^2\omega_{e}^3/6\pi\epsilon_{0}\hbar c^3 $ is the single-emitter decay rate). It can be seen that both its magnitude and phase vary only weakly across the Brillouin zone. It may thus be assumed constant around small transverse momenta of the incident light, which can be used to simplify calculations of the photon dynamics, as we shall discuss in the next section. Physically, this approximation corresponds to a short range on-site interaction between the dressed excitations. A similar diagrammatic approach for describing the propagation and scattering of polaritons in a Bose-Einstein condensate was developed in Refs.~\cite{camacho-guardian_polariton_2020,camacho-guardian_strong_2022}.

\section{Two-photon dynamics}
Having solved the scattering problem of two dressed emitters, we are now ready to analyze the non-linear two-photon dynamics of the emitter array using our Green's function formalism. To this end, we need a connection between the Green's functions and the wave function of the photons. The time-dependent Green's function yields the evolution of an $ n $-particle wave function $ \Psi^{(n)}(\vec{r}^{(n)}, t') $ of $ n $ positions $ \vec{r}^{(n)} = (\vec{r}_{1}, \dots, \vec{r}_{n}) $ to a later time $ t > t' $ as \cite{bruus_manybody_2003}
\begin{align}\label{eq:timeevol}
	\begin{split}
		\Psi^{(n)}&(\vec{r}^{(n)}, t)\\
		& = \int\dd^{3n}r^{(n)\prime}G^{(n)}(\vec{r}^{(n)}, \vec{r}^{(n)\prime}, t - t')\Psi^{(n)}(\vec{r}^{(n)\prime}, t').
	\end{split}
\end{align}
Equivalently, the Fourier transform, $ G^{(n)}(\vec{k}^{(n)}, \vec{k}^{(n)\prime}, \omega) $, of the $ n $-particle time-ordered Green's function $ G^{(n)}(\vec{r}^{(n)}, \vec{r}^{(n)\prime}, t - t') $ describes the evolution from some initial momenta $ \vec{k}^{(n)\prime} $ to some outgoing momenta $ \vec{k}^{(n)} $ at a given total energy $ \hbar\omega $. In the present case, we thus need the two-photon Green's function shown diagrammatically in \cref{fig:diagrams2}b, which consists of a term describing free propagation and an interaction term. Two incident photons are converted to dressed emitter excitations, which can propagate within the array, interact locally at a given site, and are subsequently emitted in the outgoing channel. 

Using the scattering matrix, obtained in \cref{eq:Tsol}, we can now analyze the correlated scattering dynamics of incident light as determined by the two-photon Green's function shown in \cref{fig:diagrams2}b (see \cref{app:2phGF}). It is used here to propagate an incident uncorrelated single-mode two-photon state $ \Psi_{\tu{in}}^{(2)}(\vec{k}_{1}\nu_{1}, \vec{k}_{2}\nu_{2}) = \Psi_{\tu{in}}^{(1)}(\vec{k}_{1}, \nu_{1})\Psi_{\tu{in}}^{(1)}(\vec{k}_{2}, \nu_{2}) $. To this end, we evaluate the transverse integral in \cref{eq:timeevol} in momentum space ($ \vec{k}_{\perp} $) but choose a real-space representation along the $ z $-axis, orthogonal to the emitter array. The only nontrivial contribution to the integral stems from the polarization component, $ \psi_{\tu{in}}^{(1)}(\vec{k}) = \sum_\nu \unit{e}_{+}^{\dagger}\unit{e}_{\vec{k}\nu}\Psi_{\tu{in}}^{(1)}(\vec{k},\nu) $, that couples to the emitters, which we choose here to be right-circularly polarized, as mentioned above. 

\begin{figure}
	\centering
	\includegraphics[width=\columnwidth]{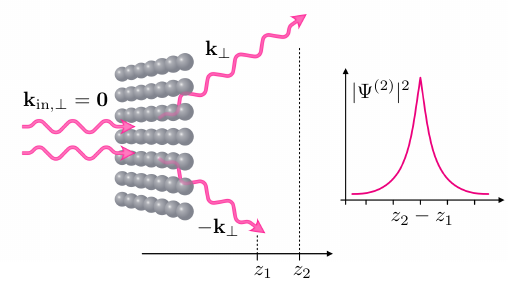}
	\caption{\label{fig:schematic} Photon-photon scattering leads to the correlated emission of photon pairs, whose localization length along the propagation direction of the incident light is determined by the collective decay rate of the array as illustrated in the figure.}
\end{figure}

While the general solution is given in \cref{app:2phwavefunction}, we discuss here a simplified expression for normally incident light, $ \psi_{\tu{in}}^{(1)}(\vec{k}) = \delta(\vec{k}_{\perp})\delta(k_{z} - k) $, which is on resonance with the zero transverse momentum collective transition of a subwavelength array. That is, we take the energy of each of the two incoming photons to be $ \omega = ck = \omega_{e} + \tilde{\Delta}_{\vec{0}} $. In this case, taking the limit $ t \rightarrow \infty $ such that all transient dynamics have died away, one obtains a compact expression for the scattered photons
\begin{widetext}
	\begin{align}
		\begin{split}\label{eq:twoPhotonWFPlane}
			\psi^{(2)}(\vec{k}_{\perp,1},\vec{k}_{\perp,2}, z_{1}, z_{2}) =&\ 2\delta(\vec{k}_{\perp,1})\delta(\vec{k}_{\perp,2}) (e^{ikz_{1}} - e^{ik|z_{1}|})(e^{ikz_{2}} - e^{ik|z_{2}|})\\
			& + \frac{a^2}{2\pi^2}\delta(\vec{k}_{\perp,1} + \vec{k}_{\perp,2})\T(\vec{0}, \Omega)\frac{\tilde{\Gamma}_{\vec{k}_{\perp, 1}}\tilde{\Gamma}_{\vec{k}_{\perp, 2}}/\tilde{\Gamma}_{\vec{0}}^2}{2\tilde{\Delta}_{\vec{0}} - \tilde{\Delta}_{\vec{k}_{\perp, 1}} - \tilde{\Delta}_{\vec{k}_{\perp, 2}} + i(\tilde{\Gamma}_{\vec{k}_{\perp, 1}} + \tilde{\Gamma}_{\vec{k}_{\perp, 2}})} e^{ik_{z}(|z_{1}| + |z_{2}|)}\\
			& \times \left(\theta(d_{2} - d_{1})e^{\left(i(\tilde{\Delta}_{\vec{0}} - \tilde{\Delta}_{\vec{k}_{\perp, 1}}) - \tilde{\Gamma}_{\vec{k}_{\perp, 1}}\right)(d_{2} - d_{1})/c} + 1 \leftrightarrow 2\right)
		\end{split}
	\end{align}
\end{widetext}
away from the emitter array, where the evanescent field can be neglected. Here, the total energy is $ \Omega = 2\omega $ and we have removed an overall phase of $ e^{-i\Omega t} $. Furthermore, $ d_{i} = \frac{k}{k_{z, i}}|z_{i}| $, with the longitudinal momentum of each photon given by $ k_{z, i} = \sqrt{k^2 - k_{\perp, i}^2} $, is the distance that a photon has propagated from the array for a given $ z_{i} $ and transverse momentum $ \vec{k}_{\perp, i} $, see \cref{fig:schematic}. This expression lets us quantify the essential mechanisms of the correlated photon interaction with the emitter array. The first line is non-zero only for $ z < 0 $ and accounts for the perfect linear reflection of the incident photons, while the remaining terms describe the nonlinear contribution to the induced photon field that is emitted symmetrically from the array in the positive and negative $ z $-direction, with an emission pattern that remarkably is given by a simple Lorentzian factor of the collective energies. This term shows that the transmitted light solely stems from the effective interaction between the emitters, with a transmission amplitude that is proportional to the $ \T $-matrix of the dressed emitter excitations. 

While the linear response to single photons preserves the transverse momentum $ \vec{k}_{\perp} = \vec{0} $, due to the suppression of Bragg scattering in a subwavelength array, the nonlinearity opens a scattering channel and facilitates the correlated emission of photon-pairs with transverse momenta $ \vec{k}_{\perp, 1} = -\vec{k}_{\perp, 2} $. In addition to the transverse correlation, the photons are emitted in spatially localized pairs along the longitudinal $ z $-direction, as described by the exponential factor $ \theta(d_{i} - d_{j})e^{-\tilde{\Gamma}_{\vec{k}_{\perp, j}}(d_{i} - d_{j})/c} $. Here, $ d_{i}/c $ is the propagation time of the $ i $'th photon since its emission from the array, and accordingly, due to the step function $ \theta(d_{i} - d_{j}) $, it is the decay rate $ \tilde{\Gamma}_{\vec{k}_{\perp, j}} $ of the last emitted photon that defines the localization length of the photon pair (see \cref{fig:schematic} for the case of normal incidence photons). This is intuitively clear as the second photon simply decays according to the single-excitation lifetime of the collective mode corresponding to its momentum. This exponential localization of two photons represents a three-dimensional free-space analogue of so-called photon bound states that appear in waveguide-QED settings, where they arise from the interaction of a single quantum emitter with photons that propagate one-dimensionally in a single transverse mode \cite{shen_strongly_2007,mahmoodian_dynamics_2020}. In fact, transmission exclusively arises from such bound states that are symmetrically emitted to either the same or opposite sides of the array. 

\begin{figure}
	\centering
	\includegraphics[width=\columnwidth]{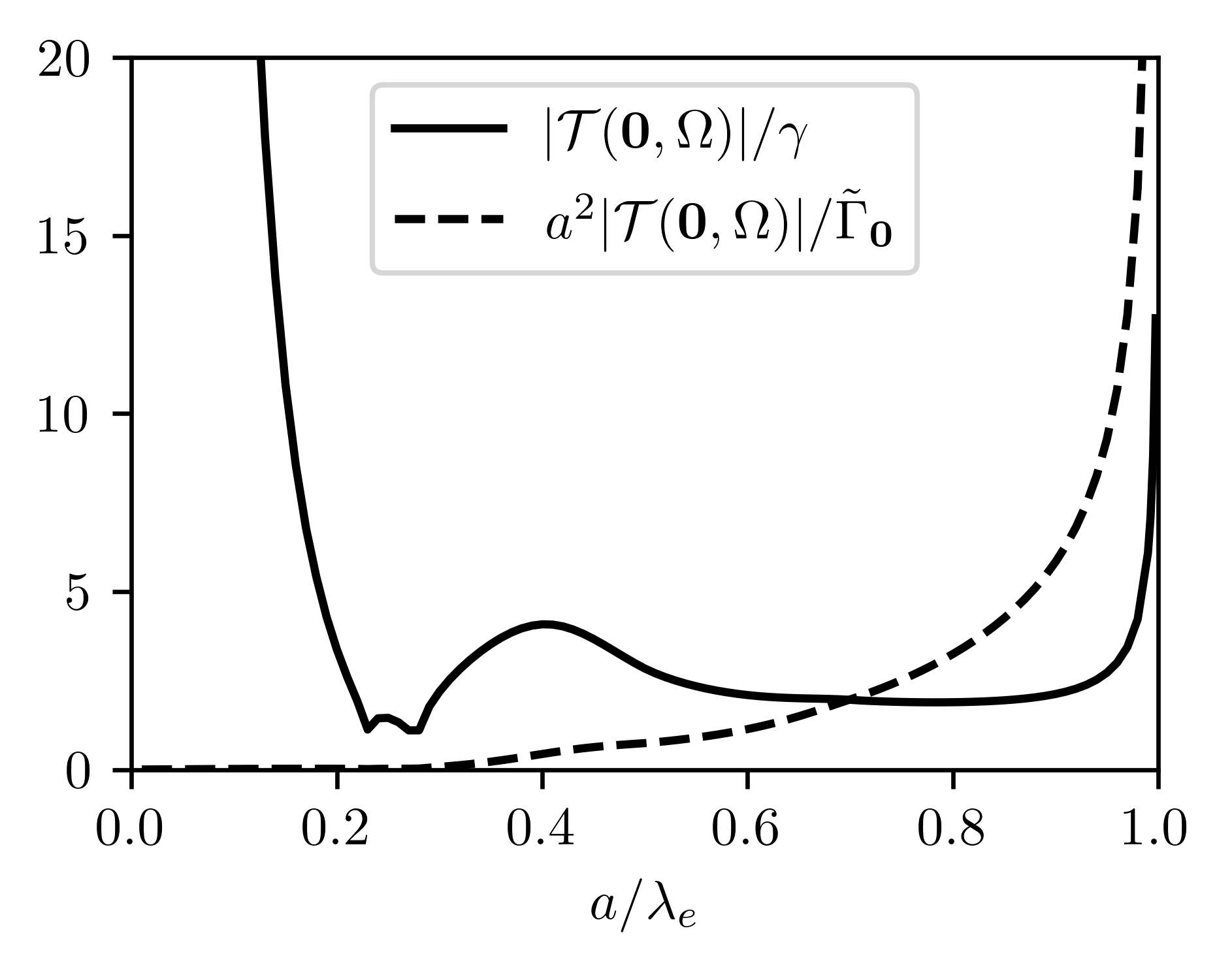}
	\caption{\label{fig:T_vs_a} The absolute value of the $ \T $-matrix in units of the single-emitter decay rate $ \gamma $, as well as $ a^2|\T(\vec{0}, \Omega)|/\tilde{\Gamma}_{\vec{0}} $, are plotted at resonant driving, $ \Omega = 2(\omega_{e} + \tilde{\Delta}_{\vec{0}}) $, for different lattice constants $ a $. The calculations show that while $ |\T(\vec{0}, \Omega)| $ increases sharply near $ a \sim 0 $ and $ a \sim \lambda_{e} $, the additional factors of $ a $ in $ a^2|\T(\vec{0}, \Omega)|/\tilde{\Gamma}_{\vec{0}} \propto a^4|\T(\vec{0}, \Omega)| $ suppress the divergence at small $ a $, but still allows for a strong nonlinearity when the emitters are resonantly spaced.}
\end{figure}

One can make the above analogy more explicit by considering a transversally localized incident photon mode and analyzing the outgoing photon flux in that mode. To be specific, we consider a Gaussian incident mode that can be well described within the paraxial approximation by a single-photon amplitude $ \psi_{\tu{in}}^{(1)}(\vec{k}) = \sqrt{2\pi}w_{0}e^{-w_{0}^2k_{\perp}^2/4}e^{ik_{z}z} $. We further assume that the spatial beam waist $ w_{0} $ is sufficiently large to confine the mode function to small transverse momenta, such that we can approximate $ \Sigma(\vec{k}_{\perp}) \approx \Sigma(\vec{0}) $ and $ \T(\vec{k}_{\perp}) \approx \T(\vec{0}) $. Even for fairly small values of $ w_{0} $, this remains a good approximation due to the weak momentum dependence of the $ \T $-matrix, illustrated in \cref{fig:T_matrix}. With these simplifications, we can project the outgoing photon state onto the Gaussian mode and obtain for the transmitted two-photon probability ($ z_{1}, z_{2} > 0 $) in the Gaussian mode (see \cref{app:2phwavefunction})
\begin{align}\label{eq:twoPhotonProp}
	P_{2}(z_{1}, z_{2}) = \frac{a^4}{w_{0}^4}\frac{|\T(\vec{0}, \Omega)|^2}{\pi^2\tilde{\Gamma}_{\vec{0}}^2} e^{ -2 \tilde{\Gamma}_{\vec{0}}|z_{2} - z_{1}|/c},
\end{align}
where we have again assumed resonant driving with $ \omega = \omega_{e} + \tilde{\Delta}_{\vec{0}} $. Hence, photons can only be transmitted in the form of an exponentially localized bound state, defined by the collective decay rate $ \tilde{\Gamma}_{\vec{0}} $. For $ a < \lambda_{e} $, $ \tilde{\Gamma}_{\vec{0}} = \frac{3\lambda_{e}^2}{4\pi a^2}\gamma \propto a^{-2} $, such that the bound state tightens with decreasing lattice spacing as the collective decay rate increases, while the overall flux of correlated photon pairs scales as $ \sim a^8|\T(\vec{0}, \Omega)|^2 $. The $ \T $-matrix sharply increases around $ a \sim 0 $ and $ a \sim \lambda $, resulting in a strong increase of photon-pair transmission as the lattice spacing approaches the transition wavelength of the quantum emitters. In \cref{fig:T_vs_a} we plot the absolute value of the $ \T $-matrix as well as $ a^2|\T(\vec{0}, \Omega)|/\tilde{\Gamma}_{\vec{0}} \propto a^4|\T(\vec{0}, \Omega)| $, which is the $ a $-dependent factor that enters our expressions. While the $ \T $-matrix diverges at both $ a = 0 $, due to the emitters coalescing in a single point, and at $ a = \lambda_{e} $, where they are at a resonant distance to each other, the full $ a $-dependent factor vanishes for small a. Thus, nonlinear effects in the emitted light are suppressed for tight lattices, but grow strong for resonantly distributed emitters.

\begin{figure}
	\centering
	\includegraphics[width=\columnwidth]{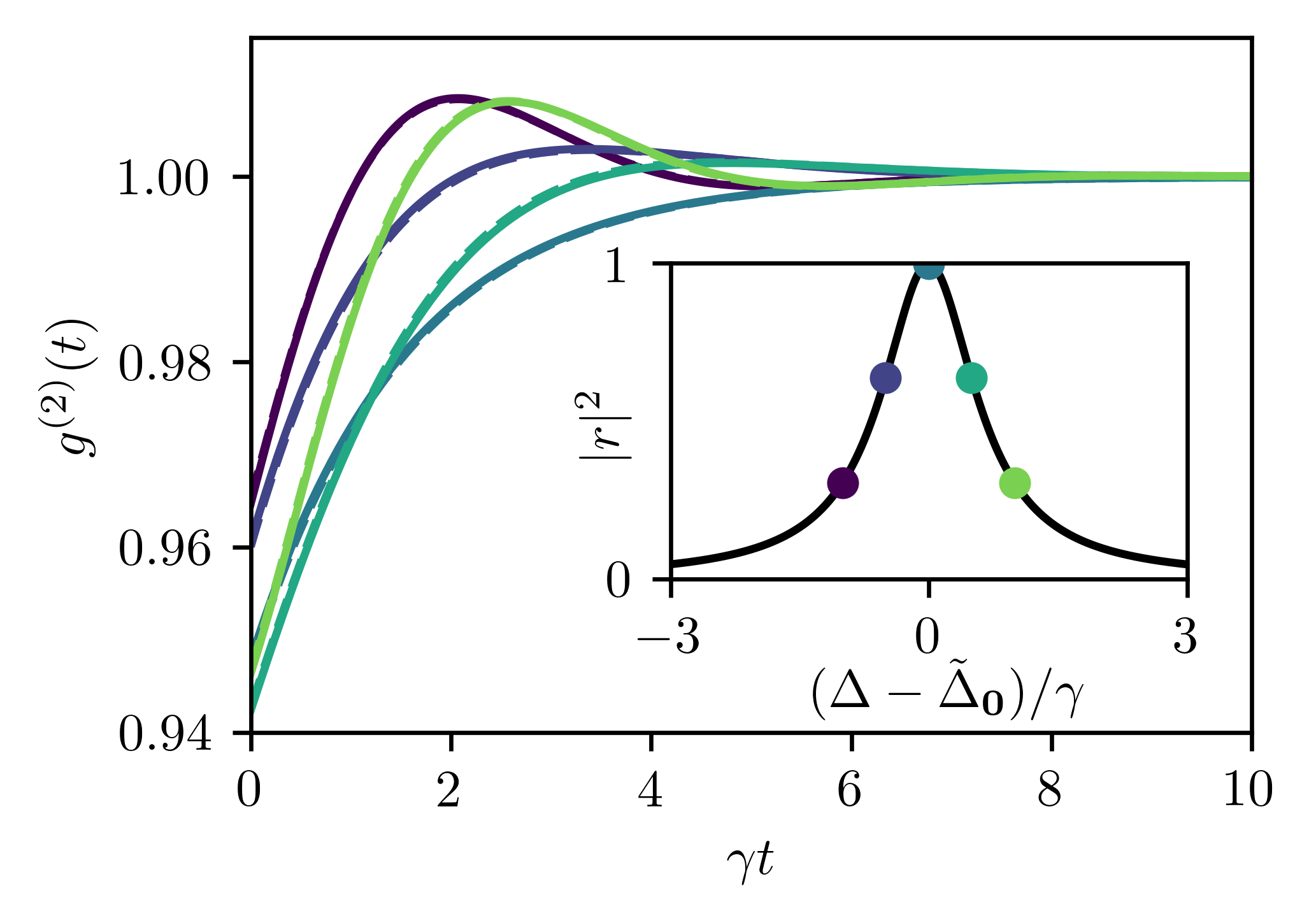}
	\caption{\label{fig:compare_g2} Two-photon correlation function $ g^{(2)}(t) $ for an incident Gaussian beam with $ w_{0} = 2.5\lambda_{e} $, $ a = 0.6\lambda_{e} $ and different detunings $ \Delta $, indicated by the colored dots the inset. The $ N \rightarrow \infty $ analytical result \cref{eq:g2Gauss} (solid lines) agrees well with the numerical simulation of a $ 15 \times 15 $ array (dashed lines).}
\end{figure}

\begin{figure}
	\centering
	\includegraphics[width=\columnwidth]{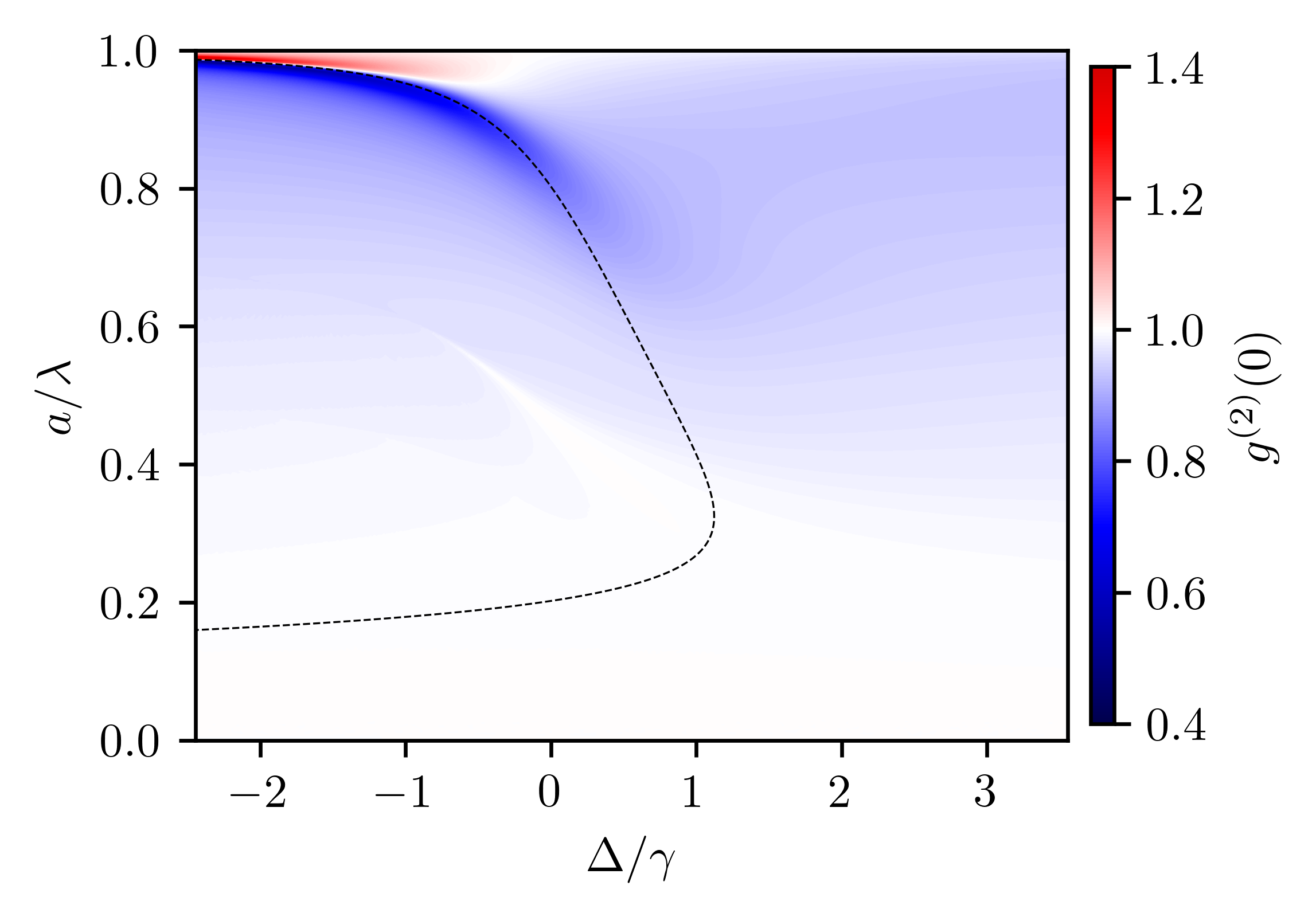}
	\caption{\label{fig:g2zero} Equal-time two-photon correlations $ g^{(2)}(0) $ as a function of the detuning $ \Delta $ and the lattice spacing $ a $. Strong antibunching occurs along the reflection resonance (white dashed line) near $ a \sim 1 $, consistent with the $ \T $-matrix shown in \cref{fig:T_vs_a}.}
\end{figure}

While the emerging bound states lead to strong photon bunching in the transmitted light, the interference with the linear response can likewise cause antibunching of reflected photons. Intuitively, we in fact expect it when driving on the reflection resonance, as independently scattered photons are near-perfectly reflected, and so any transmission of correlated photon pairs will reduce reflection, resulting in some degree of antibunching in the reflected light. An analogous calculation for the reflected light of an incident Gaussian mode yields the following simple expression for the two-photon temporal correlation function (see \cref{app:2phwavefunction})
\begin{align}\label{eq:g2Gauss}
	g^{(2)}(t) = \left|1 + \frac{a^2}{2\pi w_{0}^2}e^{(i(\Delta - \tilde{\Delta}_{\vec{0}}) - \tilde{\Gamma}_{\vec{0}})t}\frac{\T(\vec{0}, \Omega)}{\Delta - \tilde{\Delta}_{\vec{0}} + i\tilde{\Gamma}_{\vec{0}}}\right|^2
\end{align}
where $ \Delta = \omega - \omega_{e} $ is the detuning of the incident light. We see clearly how the reflected light correlations are determined by the geometric factor $ a^2/w_{0}^2 $ inversely proportional to the number of illuminated emitters, the emitter excitation scattering matrix, and the collective energies. In \cref{fig:compare_g2} we show the temporal correlations, $ g^{(2)}(t) $, for different laser detunings across the reflection resonance of an array with $ a = 0.6\lambda_{e} $, comparing the above $ N \rightarrow \infty $ analytical result with a numerical master equation simulation of a $ 15 \times 15 $ lattice using \cref{eq:in_out,eq:Heff,eq:Leff}. Despite the small lattice size and a relatively narrow beam waist of $ w_{0} = 2.5\lambda_{e} $, the linear response (inset of \cref{fig:compare_g2}) matches that of an infinite lattice, which confirms the applicability of the paraxial approximation. More remarkably, the derived expression \cref{eq:g2Gauss} for the photon-photon correlations in the infinite array limit can be seen to be an excellent approximation of the finite system already at this lattice size. This level of agreement can be understood from the fact while our analytical results concern an infinite array, the driving beam is still taken to be finite. The emitters beyond the finite illuminated area are not expected to affect the dynamics and so the infinite array becomes an excellent approximation to the finite case, which is exactly what can be seen in \cref{fig:compare_g2}. The Green's function approach thus yields an accurate description and offers an intuitive understanding of nonlinear optical processes. While instantaneous correlations, $ g^{(2)}(0) $, remains largely unaffected by the detuning, $ \Delta $, of the driving field, it has a significant effect on the temporal form of photon correlations, since it determines the relative phase between two-photon bound state and the linear contribution to the reflected light. As shown in \cref{fig:g2zero}, the degree of antibunching is largest around the reflection resonance $ \Delta = \tilde{\Delta}_{\vec{0}} $ (dashed line, as calculated by \cref{eq:selfenergy,eq:refl_ampl}). Contrary to small systems of only few emitters \cite{williamson_superatom_2020,cidrim_photon_2020}, deep subwavelength lattices show weak correlations due to the geometric factor $ a^2/w_{0}^2 $, whereas photons become increasingly antibunched as the lattice spacing approaches $ \lambda_{e} $ and the $ \T $-matrix increases (see \cref{fig:T_vs_a}). While the emitted light remains largely uncorrelated for most parameters in \cref{fig:g2zero}, under these conditions it, thus, becomes possible to convert an incident classical coherent field into pair-states of bound photons (in transmission) and antibunched light (in reflection) below the threshold $ g^{(2)}(0) < 1/2 $ \cite{scully_quantum_1997} for nonclassical light.

\section{Conclusion}
In summary, we have developed a diagrammatic Green's function formalism to describe effective photon-photon interactions and the generation of correlated states of light in two-dimensional arrays of two-level quantum emitters. Emitter excitations are treated as hardcore bosons in order to analyze their dynamics in terms of lattice polaritons, formed by hybridizing the two-dimensional collective excitations of the emitter array and the three-dimensional free-space electromagnetic field. Using Green's functions to study their interactions, we have obtained simple expressions for the optical response of the array and emerging photon-photon correlations in terms of the scattering matrix for interactions between photon-dressed emitter excitations. The derivations show that correlations can be traced back to the formation of a two-photon bound state, whose size and contribution to the total state of the scattered light feature a simple dependence on the collective linewidth and the geometry of the array. 

While we have focused here on the basic setup of a single planar square lattice of two-level quantum emitters, the outlined formalism can be readily extended to explore nonlinear processes arising from different types of emitter interactions \cite{srakaew_subwavelength_2023,moreno-cardoner_quantum_2021,zhang_photonphoton_2022,iversen_selfordering_2022}, multi-level atoms \cite{masson_dicke_2024} and more complex geometries \cite{parmee_spontaneous_2022,cech_dispersionless_2023,scheil_optical_2023,ben-maimon_quantum_2024,sierra_dicke_2022,rubies-bigorda_superradiance_2022}. Different level structures can be achieved by introducing a different interaction between the bosonized excitations instead of the hardcore repulsion used in this work. Furthermore, emitters with imperfect saturability, such as superconducting qubits, where only a finite anharmonicity separates the qubit Hilbert space from higher levels \cite{blais_circuit_2021}, can also be accurately included in our approach by taking the on-site repulsion $ U $ to be finite. As we have shown, the discussed diagrammatic formalism provides an accurate yet simple description that offers intuitive insights into the basic mechanisms for optical nonlinearities and effective photon-photon interactions, which could become useful in exploring applications of nonlinear processes in emitter arrays \cite{patti_controlling_2021,moreno-cardoner_quantum_2021,zhang_photonphoton_2022}. Importantly, it may be extended beyond the reach of numerical simulations by adapting established methods \cite{fetter_quantum_1971,bruus_manybody_2003,abrikosov_methods_1975} for a systematic inclusion of the many-body effects arising from multiple interacting photons and dressed excitations. A particularly interesting line of study would be the development of an effective low-energy theory from the two-photon propagator to describe the interacting photons, discerning the photon-photon potential, and the range of their interaction.

\section{Acknowledgments}
We thank Ephraim Shahmoon, Lida Zhang, and Klaus Mølmer for valuable discussions. This work was supported by the Austrian Science Fund (Grant No. 10.55776/COE1) and the European Union (NextGenerationEU) and through the Horizon Europe ERC synergy grant SuperWave (grant no. 101071882).

\appendix
\section{The self-energy}\label{app:selfenergy}
The free-space electromagnetic dyadic Green's function can be written as
\begin{align}\label{eq:GF_contFT}
	\vec{G}(\vec{k}_{\perp}, z, \omega) = \frac{i}{2k_{z}}\vec{Q}_{\sgn(z)}e^{ik_{z}|z|},
\end{align}
where $ \vec{Q}_{\pm} = 1 - \unit{k}\unit{k}^{\dagger} $ is the projector onto the space of polarization vectors that is orthogonal to the wave vector $ \vec{k} = (\vec{k}_{\perp}, \pm \sqrt{k^2 - k_{\perp}^2}) $ of the forward and backward propagating light.  Changing the integration variable in \cref{eq:selfE_x} to $ \omega_{\vec{k}} $, we have
\begin{align}
	\begin{split}
		\Sigma(\vec{k}_{\perp},\omega) =&\ -i\frac{\mu_{0}}{2\pi\hbar a^2}\sum_{\vec{q}_{m}}\int_{0}^{\infty}\dd\omega_{\vec{k}}\omega_{\vec{k}}^2\\
		&\hspace{20pt} \times \frac{\vec{d}^{\dagger}\tu{Im}[\vec{G}(\vec{k}_{\perp} + \vec{q}_{m}, 0, \omega_{\vec{k}})]\vec{d}}{\omega - \omega_{\vec{k}} + i\eta}.
	\end{split}
\end{align}
Using 
$\vec{G}^{*}(\vec{r}_{m}, \vec{r}_{n}, \omega) = \vec{G}(\vec{r}_{m}, \vec{r}_{n}, -\omega^{*})$ and 
$ \lim_{|\omega| \rightarrow \infty}\omega^2\vec{G}(\vec{r}_{m}, \vec{r}_{n}, \omega) = -\delta(\vec{r}_{m} - \vec{r}_{n})$ \cite{buhmann_dispersion_2012},
one can evaluate the integral 
\begin{align}\label{eq:selfenergy}
	\Sigma(\vec{k}_{\perp}, \omega) = -\frac{\mu_{0}\omega_{e}^2}{\hbar a^2}\sum_{\vec{q}_{m}}\vec{d}^{\dagger}\vec{G}(\vec{k}_{\perp} + \vec{q}_{m}, 0, \omega)\vec{d},
\end{align}
and obtain \cref{eq:selfE_x2} from $ \frac{1}{a^2}\sum_{\vec{q}_{m}}\vec{G}(\vec{k}_{\perp} + \vec{q}_{m}, 0, \omega) = N^{-1}\sum_{m, n}e^{i\vec{k}\cdot(\vec{r}_{m} - \vec{r}_{n})}\vec{G}(\vec{r}_{m}, \vec{r}_{n}, \omega) $.

\section{Single-photon transmission and reflection}\label{app:transrefl}
The single-photon amplitude evolves as
\begin{align}
	\begin{split}
		\Psi^{(1)}(\vec{k}\nu, t) =&\ i\int\frac{\dd \omega}{2\pi}e^{-i\omega t}\\
		& \times \sum_{\nu'}\int\frac{\dd^3k'}{(2\pi)^3}G_{\bar{\gamma}}^{(1)}(\vec{k}\nu, \vec{k}'\nu',\omega)\Psi_{\tu{in}}^{(1)}(\vec{k}'\nu').
	\end{split}
\end{align}
It is convenient to split $ \Psi^{(1)}(\vec{k}\nu, t) = \Psi_{0}^{(1)}(\vec{k}\nu, t) + \Psi_{\tu{sc}}^{(1)}(\vec{k}\nu, t) $ into a freely-evolving part, $\Psi_{0}^{(1)}$, and an induced component, $\Psi_{\tu{sc}}^{(1)}$, corresponding to the first and second term in \cref{eq:GdressedPhot}. The former gives $ \Psi_{0}^{(1)}(\vec{k}\nu, t) = e^{-i\omega_{\vec{k}}t}\Psi^{(1)}_{\tu{in}}(\vec{k}\nu) $. Only the projection, $ \psi^{(1)}$, onto the polarization vector of the emitter transition is affected by the second term in \cref{eq:GdressedPhot}. Here we choose right-circular polarization such that $ \psi^{(1)}(\vec{k},t) = \sum_{\nu}\unit{e}_{+}^{\dagger}\unit{e}_{\vec{k}\nu}\Psi^{(1)}(\vec{k}, \nu,t) $, as in \cref{eq:twoPhotonWFPlane}. Using the second term in \cref{eq:GdressedPhot}, the amplitude of the scattered light follows as
\begin{widetext}
\begin{align}
	\begin{split}\label{eq:Psi1_sc_preCalc}
		\psi_{\tu{sc}}^{(1)}(\vec{k}, t) &= i\frac{c\tilde{\Gamma}_{\vec{0}}}{\omega_{e}}\sqrt{\omega_{\vec{k}}}\sum_{\nu}\unit{e}_{+}^{\dagger}\unit{e}_{\vec{k}\nu}\unit{e}_{\vec{k}\nu}^{\dagger}\unit{e}_{+}\int\frac{\dd \omega}{2\pi}e^{-i\omega t}\int\frac{\dd^3k'}{2\pi}\delta_{\tu{BZ}}(\vec{k}_{\perp} - \vec{k}_{\perp}')\\
		&\hspace{50pt} \times \frac{1}{\omega + i\eta - \omega_{\vec{k}}}\frac{1}{\omega + i\eta - \omega_{e} - \Sigma(\vec{k}_{\perp}, \omega_{e})}\frac{1}{\omega + i\eta - \omega_{\vec{k}'}}\sqrt{\omega_{\vec{k}'}}\psi_{\tu{in}}^{(1)}(\vec{k}'),
	\end{split}
\end{align}
where we have used the periodicity of the self-energy in the reciprocal lattice momentum to evaluate $ \Sigma $ at the outgoing transverse momentum. Complex contour integration in $ \omega $ gives
\begin{align}
	\begin{split}
		\psi_{\tu{sc}}^{(1)}(\vec{k}, t) &= -\frac{c\tilde{\Gamma}_{\vec{0}}}{\omega_{e}}\sqrt{\omega_{\vec{k}}}\unit{e}_{+}^{\dagger}\vec{Q}\unit{e}_{+}\int\frac{\dd^3k'}{2\pi} \frac{\delta_{\tu{BZ}}(\vec{k}_{\perp} - \vec{k}_{\perp}')}{\omega_{\vec{k}'} - \omega_{e} - \Sigma(\vec{k}_{\perp}, \omega_{e})}\frac{\sqrt{\omega_{\vec{k}'}}e^{-i\omega_{\vec{k}'}t}}{\omega_{\vec{k}} - \omega_{\vec{k}'} - i\eta}\psi_{\tu{in}}^{(1)}(\vec{k}'),
	\end{split}
\end{align}
where $ \vec{Q} = \sum_{\nu}\unit{e}_{\vec{k}\nu}\unit{e}_{\vec{k}\nu}^{\dagger} $. The $ k_{z}' $ integration can be carried out for a monochromatic input field with $ \omega = c|\vec{k}'| $ in the limit $ t \rightarrow \infty $ to obtain for the total field
\begin{align}
	\begin{split}\label{eq:Psi1_sc_full_FT}
		\psi^{(1)}(\vec{k}_{\perp}z) =&\ e^{ik_{z}z}\psi_{\tu{in}}^{(1)}(\vec{k}_{\perp})
		 - \frac{i\tilde{\Gamma}_{\vec{0}}}{\omega_{e}}\frac{\omega^2}{ck_{z}}\frac{\unit{e}_{+}^{\dagger}\vec{Q}\unit{e}_{+}e^{ik_{z}|z|}}{\omega - \omega_{e} - \Sigma(\vec{k}_{\perp}, \omega_{e})}\sum_{\vec{q}_{m}}\psi^{(1)}_{\tu{in}}(\vec{k}_{\perp} + \vec{q}_{m}),
	\end{split}
\end{align}
\end{widetext}
up to an irrelevant factor $ e^{-i\omega t} $ and where $ k_{z} = \sqrt{(\omega/c)^2 - k_{\perp}^2} $. Assuming near-resonant driving ($ \Delta \ll \omega_{e} $) and neglecting evanescent fields, we can identify $ \Gamma_{\vec{k}_{\perp}} = \frac{\tilde{\Gamma}_{\vec{0}}\omega_{e}a^2}{c}\tu{Re}\left[\frac{1}{k_{z}}\unit{e}_{+}^{\dagger}\vec{Q}\unit{e}_{+}\right] = \frac{\mu_{0}\omega_{e}^2}{\hbar}\tu{Im}[\vec{d}^{\dagger}\vec{G}\vec{d}] $, which is related to the collective linewidth via $ \tilde{\Gamma}_{\vec{k}_{\perp}} = \sum_{\vec{q}_{m}}\Gamma_{\vec{k}_{\perp} + \vec{q}_{m}} $. The total amplitude
\begin{align}\label{eq:Psi1_sc_full_FT_small_Delta}
	\begin{split}
		\psi^{(1)}(\vec{k}_{\perp}z) =&\ e^{ik_{z}z}\psi_{\tu{in}}^{(1)}(\vec{k}_{\perp}) \\
		&- \frac{i\Gamma_{\vec{k}_{\perp}}e^{ik_{z}|z|}}{\Delta - \tilde{\Delta}_{\vec{k}_{\perp}} + i\tilde{\Gamma}_{\vec{k}_{\perp}}}\sum_{\vec{q}_{m}}\psi_{\tu{in}}^{(1)}(\vec{k}_{\perp} + \vec{q}_{m}).
	\end{split}
\end{align}
thus, reduces to a sum of the incident wave and Bragg-scattered components that are symmetrically emitted from the array. When Bragg scattering is suppressed at subwavelength lattice spacing, the sum only contains $ \vec{q}_{0} = \vec{0} $ and $ \Gamma_{\vec{k}_{\perp}} = \tilde{\Gamma}_{\vec{k}_{\perp}} $, such that one recovers the known expressions 
\begin{align}\label{eq:refl_ampl}
	r = -\frac{i\tilde{\Gamma}_{\vec{k}_{\perp}}}{\Delta - \tilde{\Delta}_{\vec{k}_{\perp}} + i\tilde{\Gamma}_{\vec{k}_{\perp}}},
\end{align}
and $ t = 1 - r $ for the reflection and transmission coefficient, as readily obtained from the input-output formalism of \cref{eq:in_out,eq:Heff,eq:Leff} \cite{shahmoon_cooperative_2017}.

\section{The two-photon Green's function} \label{app:2phGF}
The two-photon propagator can be written as
\begin{widetext}
	\begin{align}
		\begin{split}\label{eq:twophotonGF}
			G_{\bar{\gamma}}^{(2)}(\vec{k}_{1}\nu_{1}\omega_{1}, &\vec{k}_{2}\nu_{2}\omega_{2}; \vec{k}_{1}'\nu_{1}'\omega_{1}', \vec{k}_{2}'\nu_{2}'\omega_{2}')\\
			=& -(2\pi)^2\delta(\omega_{1} - \omega_{1}')\delta(\omega_{2} - \omega_{2}') G_{\bar{\gamma}}^{(1)}(\vec{k}_{1}\nu_{1}, \vec{k}_{1}'\nu_{1}'\omega_{1})G_{\bar{\gamma}}^{(1)}(\vec{k}_{2}\nu_{2}, \vec{k}_{2}'\nu_{2}'\omega_{2})\\
			& - (2\pi)^2\delta(\omega_{1} - \omega_{2}')\delta(\omega_{2} - \omega_{1}')  G_{\bar{\gamma}}^{(1)}(\vec{k}_{1}\nu_{1}, \vec{k}_{2}'\nu_{2}'\omega_{1})G_{\bar{\gamma}}^{(1)}(\vec{k}_{2}\nu_{2}, \vec{k}_{1}'\nu_{1}'\omega_{2})\\
			& - (2\pi)^3\delta(\omega_{1} + \omega_{2} - \omega_{1}' - \omega_{2}')\delta_{\tu{BZ}}(\vec{k}_{\perp, 1} + \vec{k}_{\perp, 2} - \vec{k}_{\perp, 1}' - \vec{k}_{\perp, 2}')\\
			&\hspace{10pt} \times \mathcal{G}_{\gamma}^{(1)}(\vec{k}_{1}, \omega_{1})g_{\vec{k}_{1}\nu_{1}}\mathcal{G}_{\bar{x}}^{(1)}(\vec{k}_{\perp, 1}, \omega_{1})\mathcal{G}_{\gamma}^{(1)}(\vec{k}_{2}, \omega_{2})g_{\vec{k}_{2}\nu_{2}}\mathcal{G}_{\bar{x}}^{(1)}(\vec{k}_{\perp, 2}, \omega_{2})\\
			&\hspace{10pt} \times 2ia^2 \T(\vec{k}_{\perp, 1}' + \vec{k}_{\perp, 2}', \omega_{1}' + \omega_{2}')\\
			&\hspace{10pt} \times \mathcal{G}_{\bar{x}}^{(1)}(\vec{k}_{\perp, 1}', \omega_{1}')g_{\vec{k}_{1}'\nu_{1}'}^{*}\mathcal{G}_{\gamma}^{(1)}(\vec{k}_{1}', \omega_{1}')\mathcal{G}_{\bar{x}}^{(1)}(\vec{k}_{\perp, 2}', \omega_{2}')g_{\vec{k}_{2}'\nu_{2}'}^{*}\mathcal{G}_{\gamma}^{(1)}(\vec{k}_{2}', \omega_{2}').
		\end{split}
	\end{align}
\end{widetext}
The first two terms describe the free propagation of the dressed photons, while the third pertains to photon-photon scattering. The latter describes the conversion of two incident photons into dressed excitations via absorption, their scattering according to $\T$, and the subsequent emission of the two photons from the array (see also \cref{fig:diagrams2}b). Although it contains many terms, the simple form of the $\T$-matrix and single-particle Green's functions permits to explicitly evolve monochromatic states, as we shall describe in more detail in the next section.

\section{The two-photon wave function}\label{app:2phwavefunction}
Let us now use \cref{eq:twophotonGF} to time-evolve an incident single-mode two-photon state $ \Psi_{\tu{in}}^{(2)}(\vec{k}_{1}\nu_{1}, \vec{k}_{2}\nu_{2}) = \Psi_{\tu{in}}^{(1)}(\vec{k}_{1}, \nu_{1})\Psi_{\tu{in}}^{(1)}(\vec{k}_{2}, \nu_{2}) $ according to 
\begin{align}
	\begin{split}
		&\Psi^{(2)}(\vec{k}_{1}\nu_{1}, \vec{k}_{2}\nu_{2}, t) = \int\frac{\dd\omega_{1}\dd\omega_{2}}{(2\pi)^2}e^{-i(\omega_{1} + \omega_{2})t}\\
		& \sum_{\nu_{1}', \nu_{2}'}\int\frac{\dd^3k_{1}'\dd^3k_{2}'\dd\omega_{1}'\dd\omega_{2}'}{(2\pi)^6}\Psi_{\tu{in}}^{(2)}(\vec{k}_{1}'\nu_{1}', \vec{k}_{2}'\nu_{2}')\\
		& \times G_{\bar{\gamma}}^{(2)}(\vec{k}_{1}\nu_{1}\omega_{1}, \vec{k}_{2}\nu_{2}\omega_{2}; \vec{k}_{1}'\nu_{1}'\omega_{1}', \vec{k}_{2}'\nu_{2}'\omega_{2}').
	\end{split}
\end{align}
Similar to the derivations in \cref{app:transrefl}, we can separate the amplitude into a linear part that is described by the first two lines on the right hand side of \cref{eq:twophotonGF} and a nonlinear part $ \Psi_{\tu{int}}^{(2)}$ that follows from the remaining terms. Projecting again onto the polarization component that couples to the emitters ($ \unit{e}_{+} $), we obtain
\begin{widetext}
\begin{align}
	\begin{split}\label{eq:Psi2_int_preCalc}
		\psi_{\tu{int}}^{(2)}(\vec{k}_{1}, \vec{k}_{2}, t) =& \sum_{\nu_{1},\nu_{2}} (\unit{e}_{+}^{\dagger}\unit{e}_{\vec{k}_{1}\nu_{1}})(\unit{e}_{+}^{\dagger}\unit{e}_{\vec{k}_{2}\nu_{2}}) \Psi_{\tu{int}}^{(2)}(\vec{k}_{1}\nu_{1}, \vec{k}_{2}\nu_{2}, t) \\
		=& -2ia^2\left(\frac{c\tilde{\Gamma}_{\vec{0}}}{\omega_{e}}\right)^2\sqrt{\omega_{\vec{k}_{1}}\omega_{\vec{k}_{2}}}\unit{e}_{+}^{\dagger}\vec{Q}_{1}\unit{e}_{+}\unit{e}_{+}^{\dagger}\vec{Q}_{2}\unit{e}_{+}\\
		& \times \int\frac{\dd^3k_{1}'\dd^3k_{2}'}{(2\pi)^4}\delta_{\tu{BZ}}(\vec{k}_{\perp, 1} + \vec{k}_{\perp, 2} - \vec{k}_{\perp, 1}' - \vec{k}_{\perp, 2}')\sqrt{\omega_{\vec{k}_{1}'}\omega_{\vec{k}_{2}'}}\psi_{\tu{in}}^{(2)}(\vec{k}_{1}', \vec{k}_{2}')\\
		& \times \int\frac{\dd\omega_{1}\dd\omega_{2}}{(2\pi)^2}e^{-i(\omega_{1} + \omega_{2})t}\mathcal{G}_{\gamma}^{(1)}(\vec{k}_{1}, \omega_{1})\mathcal{G}_{\gamma}^{(1)}(\vec{k}_{2}, \omega_{2})\mathcal{G}_{\bar{x}}^{(1)}(\vec{k}_{\perp, 1}, \omega_{1})\mathcal{G}_{\bar{x}}^{(1)}(\vec{k}_{\perp, 2}, \omega_{2})\T(\vec{k}_{\perp, 1} + \vec{k}_{\perp, 2}, \omega_{1} + \omega_{2})\\
		& \times \int\frac{\dd\omega_{2}'}{2\pi}\mathcal{G}_{\gamma}^{(1)}(\vec{k}_{1}', \omega_{1} + \omega_{2} - \omega_{2}')\mathcal{G}_{\bar{x}}^{(1)}(\vec{k}_{\perp, 1}', \omega_{1} + \omega_{2} - \omega_{2}')\mathcal{G}_{\gamma}^{(1)}(\vec{k}_{2}', \omega_{2}')\mathcal{G}_{\bar{x}}^{(1)}(\vec{k}_{\perp, 2}', \omega_{2}'),
	\end{split}
\end{align}
Carrying out the frequency integrals by complex contour integration and taking the limit $ t \rightarrow \infty $ for monochromatic incident fields with frequencies $ \omega_{1} $ and $ \omega_{2} $ gives
\begin{align}
	\begin{split}\label{eq:Psi2_int_full}
		\psi_{\tu{int}}^{(2)}(\vec{k}_{1}, \vec{k}_{2}) &= 2a^2\left(\frac{c\tilde{\Gamma}_{\vec{0}}}{\omega_{e}}\right)^2\sqrt{\omega_{\vec{k}_{1}}\omega_{\vec{k}_{2}}\omega_{1}\omega_{2}}\unit{e}_{+}^{\dagger}\vec{Q}_{1}\unit{e}_{+}\unit{e}_{+}^{\dagger}\vec{Q}_{2}\unit{e}_{+}\\
		&\hspace{20pt} \times \T(\vec{k}_{\perp, 1} + \vec{k}_{\perp, 2}, \omega_{1} + \omega_{2})\frac{1}{\omega_{1} + \omega_{2} + i\eta - \omega_{\vec{k}_{1}} - \omega_{\vec{k}_{2}}}\\
		&\hspace{20pt} \times \left(\frac{\omega_{1} + \omega_{2} - \omega_{\vec{k}_{1}} - \omega_{\vec{k}_{2}}}{\omega_{1} + \omega_{2} - \tilde{\omega}_{\vec{k}_{\perp, 1}} - \tilde{\omega}_{\vec{k}_{\perp, 2}}} + 1\right)\frac{1}{(\omega_{1} + \omega_{2} - \omega_{\vec{k}_{1}} - \tilde{\omega}_{\vec{k}_{\perp, 2}})(\omega_{1} + \omega_{2} - \tilde{\omega}_{\vec{k}_{\perp, 1}} - \omega_{\vec{k}_{2}})}\\
		&\hspace{20pt} \times \sum_{\vec{q}_{m}}\int\frac{\dd^2q_{\perp}}{(2\pi)^2}\frac{1}{(\omega_{1} - \tilde{\omega}_{\vec{k}_{\perp, 1}'})(\omega_{2} - \tilde{\omega}_{\vec{k}_{\perp, 2}'})}\psi_{\tu{in}}^{(2)}(\vec{k}_{\perp, 1}', \vec{k}_{\perp, 2}'),
	\end{split}
\end{align}
where $ \vec{k}_{\perp, i}' = (\vec{k}_{\perp, i} \pm \vec{q}_{\perp} + \vec{q}_{m}/2) $. Fourier transforming with respect to $ k_{z, i} $ and neglecting evanescent-field contributions finally yields for the two-photon amplitude
	\begin{align}
		\begin{split}\label{eq:Psi2_int_full_FT}
			\psi_{\tu{int}}^{(2)}(\vec{k}_{\perp, 1}z_{1}, \vec{k}_{\perp, 2}z_{2})	=&\ -2a^2\frac{\T(\vec{k}_{\perp, 1} + \vec{k}_{\perp, 2}, \omega_{1} + \omega_{2})}{\Delta_{1} + \Delta_{2} - \tilde{\Delta}_{\vec{k}_{\perp, 1}} - \tilde{\Delta}_{\vec{k}_{\perp, 2}} + i(\tilde{\Gamma}_{\vec{k}_{\perp, 1}} + \tilde{\Gamma}_{\vec{k}_{\perp, 2}})}e^{i(k_{z, 1}|z_{1}| + k_{z, 2}|z_{2}|)}\\
			&\hspace{10pt} \times \left(\theta(d_{2} - d_{1})e^{\left(i(\Delta - \tilde{\Delta}_{\vec{k}_{\perp, 1}}) - \tilde{\Gamma}_{\vec{k}_{\perp, 1}}\right)(d_{2} - d_{1})/c} + \theta(d_{1} - d_{2})e^{\left(i(\Delta - \tilde{\Delta}_{\vec{k}_{\perp, 2}}) - \tilde{\Gamma}_{\vec{k}_{\perp, 2}}\right)(d_{1} - d_{2})/c}\right)\\
			&\hspace{10pt} \times \sum_{\vec{q}_{m}}\int\frac{\dd^2q_{\perp}}{(2\pi)^2}\frac{\Gamma_{\vec{k}_{\perp, 1}}/a^2}{\Delta_{1} - \tilde{\Delta}_{\vec{k}_{\perp, 1}'} + i\tilde{\Gamma}_{\vec{k}_{\perp, 1}'}}\frac{\Gamma_{\vec{k}_{\perp, 2}}/a^2}{\Delta_{2} - \tilde{\Delta}_{\vec{k}_{\perp, 2}'} + i\tilde{\Gamma}_{\vec{k}_{\perp, 2}'}}\psi_{\tu{in}}^{(2)}(\vec{k}_{\perp, 1}', \vec{k}_{\perp, 2}'),
		\end{split}
	\end{align}
\end{widetext}
where $ k_{z, i} = \sqrt{(\omega_{i}/c)^2 - k_{\perp, i}^2} $ and $ d_{i} = \frac{k}{k_{z, i}}|z_{i}| $. Considering normal-incident plane wave fields, $ \psi_{\tu{in}}^{(1)}(\vec{k}_{i}) = \delta(\vec{k}_{\perp, i})\delta(k_{z,i} - \omega_{i}/c) $ and omitting Bragg scattering for subwavelength arrays, one arrives at \cref{eq:twoPhotonWFPlane}. Likewise, choosing a broad Gaussian beam, $ \psi_{\tu{in}}^{(1)}(\vec{k}_{i}) = \phi_{G}(\vec{k}_{i}) = \sqrt{2\pi}w_{0}e^{-w_{0}^2k_{\perp, i}^2/4}e^{ik_{z, i}z} $, and projecting $ \psi^{(2)} $ onto the same Gaussian mode
\begin{align}
	\begin{split}
		\psi_{G}^{(2)}(z_{1}, z_{2}) =& \int\frac{\dd^2k_{\perp, 1}\dd^2k_{\perp, 2}}{(2\pi)^4}\phi_{G}^{*}(\vec{k}_{\perp, 1}z_{1})\phi_{G}^{*}(\vec{k}_{\perp, 2}z_{2})\\
		&\hspace{20pt} \times \psi^{(2)}(\vec{k}_{\perp, 1}z_{1}, \vec{k}_{\perp, 2}z_{2})
	\end{split}
\end{align}
one can arrive at \cref{eq:twoPhotonProp} via
\begin{align}
	P_{2}(z_{1}, z_{2}) = \left|\psi_{G}^{(2)}(z_{1}, z_{2})\right|^2
\end{align}
by taking $ z_{1}, z_{2} > 0 $. There, the overlap between $ \psi^{(1)}(\vec{k}_{\perp, i}z_{i}) $ and $ \phi_{G}(\vec{k}_{\perp, i}z_{i}) $ is zero, when the Gaussian mode function is confined to small transverse momenta, such that we can approximate $ \Sigma(\vec{k}_{\perp}) \approx \Sigma(\vec{0}) $, resulting in the beam being fully reflected, and $ P_{2} $ only containing the nonlinearly transmitted contribution. Analogously, one can arrive at \cref{eq:g2Gauss} via
\begin{align}
	g^{(2)}(t) = \left|\frac{\psi_{G}^{(2)}(z_{2} - z_{1} = ct)}{2(\psi_{G}^{(1)})^2}\right|^2
\end{align}
Here, we exploit the fact that $ \psi_{G}^{(2)}(z_{1}, z_{2}) $ only depends on the relative distance traveled by the photons, $ |z_{2} - z_{1}| $, to replace it with relative propagation time $ ct $. This is due to the fact that $ \psi_{G}^{(1)} $ is independent of $ z $, as the $ z $-dependent phase of $ \psi^{(1)} $ cancels with that of $ \phi_{G} $. Finally, the factor of $ 2 $ in the denominator comes from the fact that the two non-interacting diagrams of \cref{fig:diagrams2}b contribute equally to the linear part of $ \psi_{G}^{(2)}(z_{1}, z_{2}) $.

%

\end{document}